\documentstyle[prl,twocolumn,aps,epsf]{revtex}

\newcommand{\beq}{\begin{equation}}
\newcommand{\eeq}{\end{equation}}
\newcommand{\bea}{\begin{eqnarray}}
\newcommand{\eea}{\end{eqnarray}}
\newcommand{\ApJ}[1]{Astrophys.~J. {\bf #1}}
\newcommand{\AandA}[1]{Astron.\ Astrophys. {\bf #1}}
\newcommand{\PRL}[1]{Phys.\ Rev.\ Lett. {\bf #1}}
\newcommand{\PR}[2]{Phys.\ Rev.\ #1 {\bf #2}}
\begin{document}
\draft
\title{
Ion structure factors and electron transport 
in dense Coulomb plasmas\thanks{\PRL{81}, No.\,24 (14 December 1998)}
}
\author{D.A.\ Baiko, A.D.\ Kaminker, A.Y.\ Potekhin, and D.G.\ Yakovlev}
\address{Ioffe Physical-Technical Institute,
      Politekhnicheskaya 26,
     194021 St.-Petersburg, Russia}
\date{Received 21 August 1998}
\maketitle

\begin{abstract}
The dynamical structure factor of a Coulomb crystal of ions
is calculated at arbitrary temperature below
the melting point taking into account multi-phonon
processes in the harmonic approximation.
In a strongly coupled Coulomb ion liquid,
the static structure factor is split into two parts,
a Bragg-diffraction-like one,
describing incipient long-range order structures,
and an inelastic part corresponding
to thermal ion density fluctuations.
It is assumed that the diffractionlike scattering
does not lead to the electron relaxation
in the liquid phase. 
This assumption, together with the inclusion
of multi-phonon processes in the crystalline phase,
eliminates large discontinuities of the transport coefficients
(jumps of the thermal and electric conductivities, 
as well as shear viscosity, reported previously)
at a melting point.
\end{abstract}


% 52.25.-b: Plasma properties
% 52.25.Fi: Transport properties
% 95. Fundamental A&A
% 95.30.Qd: MHD and plasmas
% 97.20.Rp: Faint blue stars, white dwarfs, degenerate stars, nuclei of planetary nebulae
% 97.60.Jd: Neutron stars

We consider
a strongly coupled Coulomb plasma (SCCP) of ions
immersed in a nearly uniform charge-compensating electron gas.
The ions may be disordered (liquid phase)
or arranged in a crystalline lattice.
The energetically favorable
body-centered cubic (bcc) lattice,
appears at $\Gamma > \Gamma_m \approx 172$
\cite{NNN}, where $\Gamma=(Ze)^2/(aT)$ is the
ion-coupling parameter, $T$ is the temperature, $a=(4 \pi n_i /3)^{-1/3}$,
and $n_i$ is the ion number density.

Many astrophysical objects (interiors of white dwarfs, massive stars,
and giant planets;
envelopes of neutron stars) are made of such a plasma.
Its kinetic properties
required for various applications
are determined mainly by electron-ion ($ei$) scattering.
A general framework for calculation of
these transport properties has been described in \cite{FI76}.
Numerous calculations
(e.g., \cite{Itoh84,Itoh93,Itoh94,BY95,PCY97}), done under additional
assumption of strong electron degeneracy,
predict large (a factor of 3--4) discontinuities
of the electric and thermal conductivities at the melting point.
In contrast, the thermodynamic quantities
in the liquid and solid phases, determined solely by ions,
are very similar near $\Gamma=\Gamma_m$
(e.g., \cite{PollockHansen,NNN}).
This suggests that properties of the ion system serving
as a main scatterer for electrons should vary
smoothly through the melting transition.
In this Letter, we propose a modification of the transport
theory which removes large jumps of the transport coefficients.

The differential $ei$ scattering rate in a SCCP
averaged over initial and summed over final
electron spin states $\sigma$ and $\sigma'$ is
\bea
   \Gamma({\bf p}\to{\bf p}') &=&
   {2 \pi N \over \hbar^2}
   \frac12 \sum_{\sigma\sigma'} \left| U_{{\bf q}, \sigma' \sigma} \right|^2
   {\cal S}({\bf q}, \omega),
\\
   {\cal S}({\bf q}, \omega) &=& {1 \over 2 \pi} \int^{+\infty}_{-\infty} 
   {\rm d}t \,
   e^{-i \omega t} S({\bf q},t) 
\nonumber\\ &=&
   {1 \over 2 \pi N} \int^{+\infty}_{-\infty} {\rm d}t
   \int {\rm d}{\bf x} \, {\rm d}{\bf x}' \,
   e^{i {\bf q}\cdot ({\bf x} - {\bf x}') - i \omega t}
\nonumber\\&&\times
   \left\langle \hat{\rho}^\dagger ({\bf x},t) \,
   \hat{\rho} ({\bf x}',0)  \right\rangle_T,
\eea
where 
$N$ is the total number of ions, 
{\bf p} and ${\bf p}'$ are the electron momenta
before and after scattering, respectively,
$\hbar {\bf q} = {\bf p}' - {\bf p}$,
$\hbar \omega=\epsilon'-\epsilon$ is the difference
between final and initial electron energies, and
$U_{{\bf q}, \sigma' \sigma}$ is the matrix element of the
operator of elementary $ei$ interaction.
${\cal S}({\bf q},\omega)$
is the dynamical structure factor of the plasma,
the most important quantity of the theory.
In the liquid regime, $\hat{\rho} ({\bf x},t)$ 
is the operator of the charge density in units of $Z|e|$:
$\hat{\rho} ({\bf x},t)
= {\rm \hat{n}_I} ({\bf x},t) - n_i$,
where ${\rm \hat{n}_I} ({\bf x},t)$
is the ion density operator and $n_i=n_e/Z$ 
takes account of the compensating electron background
with the electron density $n_e$.
In the solid regime,
 $\hat{\rho} ({\bf x},t)
= {\rm \hat{n}_I} ({\bf x},t) - \sum_i \delta({\bf x} - {\bf R}_i)$
(where ${\bf R}_i$ is a lattice vector), i.e.\ the operator of fluctuations
of the charge density.

Integrating over {\bf x} and ${\bf x}'$ we obtain
the structure factor of the ion density fluctuations
in the solid phase in the form
\bea
     N S_{\rm sol}({\bf q},t) &=& \left\langle \sum_{i,j}
     e^{i {\bf q}\cdot ({\bf R}_i - {\bf R}_j)}
     \left[e^{i {\bf q}\cdot {\bf u}_i(t)} -1\right]
\right.\nonumber\\&&\left.\phantom{\sum_{i,j}}   \times
     \left[e^{-i {\bf q}\cdot {\bf u}_j(0)} -1\right] \right\rangle_T,
\label{Sqt-det}
\eea
where ${\bf u}_i$
is an ion displacement from ${\bf R}_i$.
Expanding ${\bf u}_i$ in the phonon normal coordinates
and using the Weyl operator identity $e^A e^B
= e^{A+B} e^{[A,B]/2}$,
we can decompose $S_{\rm sol}({\bf q},t)$ into
the elastic (Bragg) and inelastic parts
$S_{\rm sol}({\bf q},t) = S_{\rm sol}'({\bf q}) + S_{\rm sol}''({\bf q},t)$. 
The elastic part
is easily calculated (cf.\cite{K63}):
\beq
    S_{\rm sol}'({\bf q}) = (1-e^{-W})^2 
      (2\pi)^3 n_i \sum_{\bf G} \delta({\bf q}-{\bf G}),
\eeq
where ${\bf G}$ is a reciprocal lattice vector, 
and $W=W({\bf q})$ is the Debye-Waller factor,
$\exp(-W) = \left\langle \exp(i{\bf q}\cdot{\bf u}_j) \right\rangle_T$,
\beq
    W = {\hbar\over2MN} \sum_\nu
      {({\bf q}\cdot{\bf e}_\nu)^2 \over\omega_\nu}
         \left(\bar{n}_\nu+\frac12\right).
\label{DW}
\eeq
In this case, $M$ is the ion mass,
$\nu \equiv ({\bf k},s)$,
$s=1,2,3$ enumerates phonon modes,
{\bf k} is a phonon wavevector,
${\bf e}_\nu$ the polarization vector, $\omega_\nu$
the frequency, and
$\bar{n}_\nu =
\left( e^{z_\nu}-1 \right)^{-1}$ is
the mean number of phonons, $z_\nu=\hbar \omega_\nu/T$. For 
the lattice types of interest (e.g., bcc),
$W = r_T^2 q^2 / 6$, 
where $r_T^2 = \langle {\bf u}^2\rangle_T$ 
is the mean-squared ion displacement (cf.\cite{BY95,K63}).

The Bragg scattering of electrons results in the energy band
structure of the electron Bloch states,
but does not contribute to the $ei$ collision integral
in the kinetic equation \cite{FI76}.
Indeed, this scattering occurs at the boundaries
of the Brillouin zones and translates an electron from one zone
to another. The transition requires change of the electron
energy by the value of the interband gap; thus
another particle must be involved to carry the excess energy.

Therefore only the inelastic part of the structure factor
contributes to the collision integral.
The inelastic part can be found by the same technique \cite{K63}:
\bea &&
   N S_{\rm sol}''({\bf q},t) = e^{-2W}
      \sum_{ij} e^{i{\bf q}\cdot ({\bf R}_i-{\bf R}_j)}
        \sum_{n=1}^\infty {1\over n!} 
\nonumber\\&&\,\,\times
       \left\{
           {\hbar\over2MN} \sum_\nu
      {({\bf q} \cdot {\bf e}_\nu)^2 \over \omega_\nu}
       \left[\alpha_{ij\nu} (\bar{n}_\nu+1)
                + \alpha_{ij\nu}^\ast \bar{n}_\nu
       \right] \right\}^n ,
\label{S2}
\eea
where $\alpha_{ij\nu} \equiv
\exp[i{\bf k}\cdot ({\bf R}_i - {\bf R}_j) - i \omega_\nu t]$.
The summation over $i,j$
yields the delta function which removes one
summation over {\bf k} (included in the sums over $\nu$).
Thus we have
$n$ sums over $s$ and $n-1$ sums over ${\bf k}$
in each $n$th term of Eq.\ (\ref{S2}).

Retaining the first term $n=1$, we recover
the one-phonon approximation employed in
previous works (e.g., \cite{FI76,Itoh84,Itoh93,Itoh94,BY95}).
Our point is that this approximation fails
near the melting point.
In fact, the contribution of the $n$-phonon
processes ($n$th term) at $T$ above the Debye temperature
can be estimated as
$(q r_T)^{2n}/n! \sim (k_F r_T)^{2n}/n!$,
where $k_F = (3 \pi^2 n_e)^{1/3}$ is the electron Fermi wave number,
$r_T^2 \approx u_{-2} a^2/\Gamma$,
$u_{-2}\equiv\langle \omega_p^2/\omega_\nu^2\rangle_{\rm ph}
\approx 13$ is a frequency moment
for a bcc lattice \cite{PollockHansen},
$\omega_p$ is the ion plasma frequency, and
$\langle \ldots \rangle_{\rm ph}$
denotes averaging over phonon spectrum (e.g., \cite{BY95})
in the harmonic-lattice approximation. For instance, for Fe 
plasma at
$\Gamma \approx \Gamma_m$
we obtain a typical value of
$n \sim (k_F r_T)^2\approx 3$, which is not small.

An important difference of
astrophysical Coulomb crystals from the terrestrial metals
is that the umklapp processes dominate the scattering,
because the equivalent radius of the Brillouin zone 
$q_B=(6\pi^2 n_i)^{1/3}$
is smaller than typical momentum transfers $q \sim k_F$. For $q>q_B$ 
one can approximately
replace $\sum_s ({\bf q}\cdot{\bf e}_\nu)^2 f(\omega_\nu) \to
q^2 \langle f(\omega_\nu) \rangle_{\rm ph}$ \cite{RY82}.
Then the remaining summations in Eq.\ (\ref{S2})
are done explicitly:
\bea
&&
      e^{2W} \, S_{\rm sol}''({\bf q},t) = 
\nonumber\\&&
      \exp\left[
      {\hbar q^2\over2M} \left\langle
      {\cos(\omega_\nu t) \over \omega_\nu \,
     \tanh(z_\nu/2)}
      -i\,{\sin(\omega_\nu t)
        \over \omega_\nu}\right\rangle_{\rm ph}
       \right] - 1 .
\label{S''}
\eea

The static structure factor is defined as $S({\bf q}) = 
\int_{-\infty}^{+\infty}{\cal S}({\bf q},\omega)\,{\rm d}\omega
= S({\rm q},0)$.
Thus the inelastic contribution to $S({\bf q})$ 
in the solid is given by setting $t=0$ in Eq.~(\ref{S''}),
which yields
$S_{\rm sol}''({\bf q}) = 1 - e^{-2W}$.

Let us turn to the liquid phase.
Numerical simulations by different authors show the appearance of
incipient long-range order at $\Gamma\gg1$. For 
example, Schmidt et al.\ \cite{MC} observed
a shear mode at $\Gamma > 100$ in their molecular-dynamics experiment
along with the familiar longitudinal ion plasmon. We have verified that
the spectrum of these modes can be described
by the phonon spectrum averaged over
orientations of a crystal.
Although the long-range order does not persist forever,
it may be well
preserved during typical electron scattering time.
Thus a temporary
electron band structure emerges, and an associated elastic scattering
does not contribute to the conduction (as in solid).
This is in line with Edwards's \cite{Edwards} argument that
one should deal with a local disorder ``observed'' by an electron
along its mean free path,
rather than with the global disorder.
Therefore we suggest to subtract the elastic contribution from the
total static structure factor $S_{\rm liq}(q)$ in the liquid
(e.g., \cite{Hansen73,Young91}).
Since in the liquid an electron
couples directly to the ion number density, the elastic part
must have the form \cite{K63}
\beq
    S_{\rm liq}'({\bf q}) = e^{-2W}
       (2\pi)^3 n_i \sum_{{\bf G}\neq0} \delta({\bf q}-{\bf G}).
\eeq
Then the ``inelastic'' part which determines
the transport properties is 
$S_{\rm liq}''(q)=S_{\rm liq}(q)-S_{\rm liq}'({\bf q})$.
There may be various types of periodic structures in this regime,
but they are very similar and we can
use the bcc lattice.
We have checked that the result is almost the same
for face-centered cubic (fcc) and hexagonal close-packed (hcp) lattices.

At this stage we need
to specify the matrix element of the elementary $ei$ interaction
$U_{{\bf q}, \sigma' \sigma}$.
Assuming the Coulomb potential
screened by the static polarization of ideal,
relativistic, strongly degenerate
($p \approx p' \approx  \hbar k_F$) electrons,
we obtain
\beq
   {2 \pi N \over \hbar^2} \frac12 \sum_{\sigma\sigma'}
   \left| U_{{\bf q}, \sigma' \sigma} \right|^2
   = {2 \pi N \over\hbar^2 V^2} \,
   {16 \pi^2 Z^2 e^4 \over q^4 |\varepsilon(q)|^2}
   \left(1- {\hbar^2 c^2 q^2 \over 4 \epsilon_F^2} \right),
\eeq
where $\epsilon_F =[m^2_e c^4+p_F^2 c^2]^{1/2}$
is the electron Fermi energy, $V$ is the
normalization volume, and
$\varepsilon(q)$ is the electron longitudinal
static dielectric function \cite{J62}.

The electric ($\sigma$) and thermal ($\kappa$) conductivities,
and shear viscosity ($\eta$)
can be written in the form \cite{FI76}
\beq
   \sigma={n_e e^2 \over m_e^\ast \nu_\sigma},
\quad
   \kappa = {\pi^2 T n_e\over 3 m_e^\ast \nu_\kappa},
\quad
   \eta = {n_e m^\ast_e v^2_F \over 5 \nu_\eta},
\eeq
where
$m_e^\ast = \epsilon_F/c^2$,
and
\beq
   \nu_{\sigma,\kappa,\eta} = 4 \pi m^\ast_e Z^2 e^4 \,
                         L_{\sigma,\kappa,\eta} \, n_i /(\hbar k_F)^3
\eeq
%

%                                                       FIGURE fig1
\begin{figure}[ht]
\begin{center}
\leavevmode
\epsfysize=95mm
\epsfbox{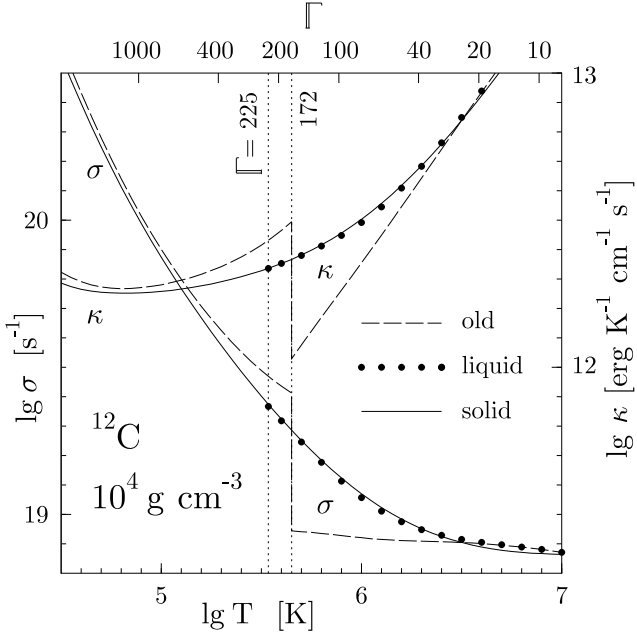}
\end{center}
\caption[ ]{
Dependence of the electron electric (left
vertical scale) and thermal (right vertical scale)
conductivities of carbon plasma at density $10^4$ g cm$^{-3}$
on temperature (lower horizontal scale) or ion coupling parameter
(upper horizontal scale).
Dashes are traditional results in ion liquid
\protect{\cite{Itoh84,PCY97}} and bcc crystal
\protect{\cite{Itoh93,BY95}} for $\Gamma_m=172$.
Solid line is the present multi-phonon calculation in solid,
extended artificially to high $T$; 
dots show the
present calculation with reduced structure
factor in liquid at $\Gamma \leq 225$. Vertical dotted lines
correspond to $\Gamma = 172$ and 225.
}
\label{fig1}
\end{figure}

\noindent
are the effective collision frequencies.
Here, the effective Coulomb logarithms are
\bea
   L_{\sigma,\kappa} &=& \int_{q_0}^{2k_F}
{{\rm d}q\,q^3 \over q^4 |\varepsilon(q)|^2} \left[1-\frac14\,
\left({\hbar q\over m_e^\ast c } \right)^2 \right] S_{\sigma,\kappa}(q),
 \\
  L_\eta &=& 3 \int_{q_0}^{2k_F}
           {{\rm d}q\,q^3 \over q^4 |\varepsilon(q)|^2}
           \left(1-{q^2 \over 4 k^2_F}\right)
\nonumber\\&&\times
     \left[1-\frac14\, \left({\hbar q\over m_e^\ast c } \right)^2 \right]
           S_\sigma(q),
\eea
$q_0=0$ for the liquid phase and $q_0=q_B$ for the solid phase \cite{BY95},
and $S_{\sigma,\kappa}(q)$ are the {\it effective\/} static structure factors.
In the liquid regime,
we approximate
 $S_{\sigma,\kappa}(q)$ by $S_{\rm liq}''(q)$ 
as described above. In the solid regime, we have
\bea
   S_{\sigma}(q) &=& {1 \over 2\pi }
    \int_{-\infty}^{+\infty} {\rm d}\omega
    \int_{-\infty}^{+\infty} {\rm d}t \,\,  { e^{-i\omega t}\,
       z \over 1- e^{-z} } \,
        S_{\rm sol}''({\bf q},t),
\\
   S_{\kappa}(q) &=& S_{\sigma}(q) +
          \left( {3 k_F^2\over q^2}-\frac12\right) \delta S_\kappa(q),
\\
   \delta S_\kappa(q) &=&  {1 \over 2 \pi }
    \int_{-\infty}^{+\infty} {\rm d}\omega
    \int_{-\infty}^{+\infty} {\rm d}t \,\, { e^{-i\omega t}\,
        z^3 \over 1-e^{-z} } \,
        S_{\rm sol}''({\bf q},t),
\eea
with $z=\hbar \omega / T$.
The integration over $\omega$
can be performed analytically.
The remaining numerical integration over $t$
is then facilitated by shifting the integration path
in complex plane: $t=t'-i \hbar/2T$, where $t'$ is real.
The final result reads:
\bea
   S_{\sigma}(q) &=& {1  \over 2 }
    \int_{-\infty}^{+\infty} { {\rm d}x \over \cosh^2 x }\,
    e^{-2W} \, K(q,T,t'),
\\
   \delta S_\kappa(q) &=&
         \int_{-\infty}^{+\infty} {\rm d}x \,
        {1-2 \sinh^2 x \over \cosh^4 x} \,
   e^{-2W} \,K(q,T,t'),
\eea
where $x=\pi \, t' T/\hbar$ and
\beq
   K(q,T,t') =\exp\left[{\hbar q^2\over2M}\left\langle
        {\cos(\omega_\nu t') \over\omega_\nu \,
          \sinh(z_\nu/2) }
         \right\rangle_{\rm ph}
        \right] - 1.
\eeq
Retaining the term $\propto q^2$ in the expansion
of $K(q,T,t)$, we recover the standard one-phonon approximation
\cite{Itoh84,Itoh93,Itoh94,BY95}.

\begin{figure}[ht]
\begin{center}
\leavevmode
\epsfysize=95mm
\epsfbox{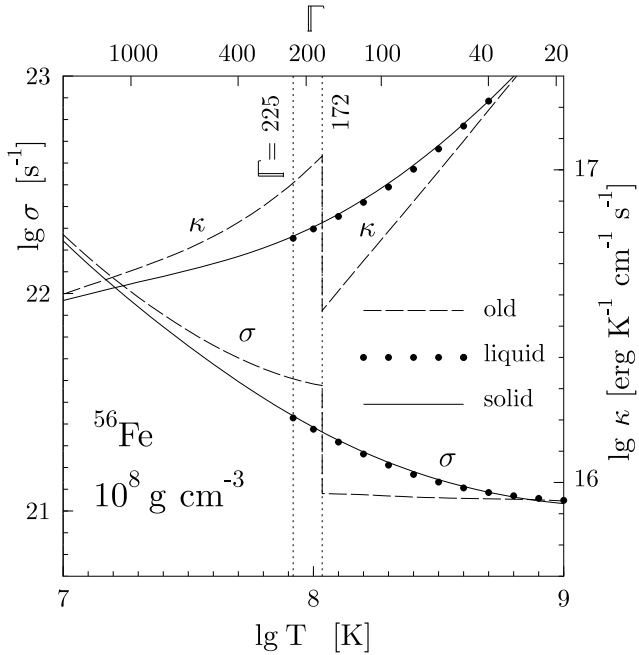}
\end{center}
\caption[ ]{
Same as in Fig.\ \protect{\ref{fig1}} but for
iron matter at density $10^8$ g cm$^{-3}$.
}
\label{fig2}
\end{figure}

Figures~\ref{fig1} and \ref{fig2} show temperature dependence of
the electric and thermal conductivities
for carbon plasma at density $10^4$ g cm$^{-3}$ and
for iron plasma at $10^8$ g cm$^{-3}$, respectively, calculated
in the Born approximation.
In spite of large differences in densities and
chemical elements, the figures are fairly similar.
Dashes show the traditional results
calculated with the full structure factor $S_{\rm liq}(q)$ in ion liquid
\cite{Itoh84,PCY97} and in the one-phonon approximation
\cite{Itoh93,BY95} for bcc crystal (notice that the results
of ref.\ \cite{BY95} for fcc crystals are in error;
actually, they are very similar to those for bcc).
One can see strong jumps of $\kappa$ and $\sigma$ at the
melting point $\Gamma_m=172$. Solid lines
are the present results in the solid phase (including
multi-phonon processes), while dots show the present
results in the liquid obtained using the analytic fits
for the static structure factor $S_{\rm liq}(q)$ at $\Gamma \leq 225$ \cite{Young91}
by subtracting the long-range correlations (see above).
For illustration (as suggested by H.E.\ DeWitt),
we have extended the improved results in liquid by
shifting artificially the melting point
to lower $T$ (to $\Gamma=225$, considering thus supercooled liquid)
and the improved results in solid by shifting the melting
to higher $T$ (lower $\Gamma$, superheated crystal).
The curves for liquid and solid ions match one another
quite well, and the jumps at the melting point actually
disappear. In a wide temperature range 
the improved curves for liquid and solid almost coincide.
We have verified that the same is true for all transport
coefficients (including shear viscosity)
in wide range of
densities for a number of chemical elements.
Thus one can observe that electron transport properties in SCCP
of ions appear to be fairly insensitive to the state
of SCCP (liquid or solid, bcc or other crystals).
This should be taken into account in various astrophysical
applications, for instance, in calculating the temperature growth
from the surface into interior of
 the isolated neutron stars important for theories
of neutron star cooling \cite{GPE83,PCY97}. The thermal
conductivity of neutron-star envelopes to be reconsidered
lies exactly in the ``sensitivity strip'' \cite{GPE83} which
strongly affects the temperature profiles and neutron-star cooling.

%                                                       FIGURE fig2

\vspace*{1cm}

We are grateful to H.E.\ DeWitt and F.J.\ Rogers for useful discussions.
This work was supported in part by RFBR (grant 96--02--16870a),
RFBR-DFG (grant 96--02--00177G),
and INTAS (grant 96--0542).


\begin{references}

\bibitem{NNN}
H. Nagara, Y.~Nagata, and T.~Nakamura,
\PR{A}{36}, 1859 (1987)

\bibitem{FI76}
E.~Flowers and N.~Itoh,
\ApJ{206}, 218 (1976)

\bibitem{Itoh84}
N.~Itoh, Y.~Kohyama, N.~Matsumoto, and M.~Seki,
\ApJ{285}, 758 (1984)

\bibitem{Itoh93}
N.~Itoh, H.~Hayashi, and Y.~Kohyama,
\ApJ{418}, 405 (1993);  {\bf 436}, 418 (E) (1994)

\bibitem{Itoh94}
   N.\ Itoh,  in {\it The Equation of State in Astrophysics},
   edited by G.~Chabrier and E.~Schatzman 
   (Cambridge University Press, Cambridge, 1994), p. 394

\bibitem{BY95}
   D.A.\ Baiko and D.G.\ Yakovlev,
   Astron.\ Lett.\ {\bf 21}, 702 (1995)
    
\bibitem{PCY97}
   A.Y.\ Potekhin, G.~Chabrier, and D.G.\ Yakovlev,
   \AandA{323}, 415 (1997)

\bibitem{PollockHansen}
E.L. Pollock and J.P. Hansen,
\PR{A}{8}, 3110 (1973)

\bibitem{K63}
C.~Kittel,
{\it Quantum Theory of Solids} 
(Wiley, New York, 1963)

\bibitem{RY82}
M.E. Raikh and D.G. Yakovlev,
Astrophys.\ Sp.\ Sci. {\bf 87}, 193 (1982)

\bibitem{MC}
P.~Schmidt, G.~Zwicknagel, P.G. Reinhard, C.~Toepffer,
\PR{E}{56}, 7310 (1997)

\bibitem{Edwards}
S.F. Edwards,
Proc.\ R.\ Soc. A {\bf 267}, 518 (1962)

\bibitem{Hansen73}
J.P. Hansen,
\PR{A}{8}, 3096 (1973)

\bibitem{Young91}
   D.A.\ Young, E.M.\ Corey, and H.E.\ DeWitt,
   \PR{A}{44}, 6508 (1991)

\bibitem{J62}
   B.\ Jancovici, Nuovo Cim.\ {\bf 25}, 428 (1962)

\bibitem{GPE83}
      E.H.\ Gudmundsson, C.J.\ Pethick, and R.I.\ Epstein,
      \ApJ{272}, 286 (1983)

\end{references}
\end{document}